\documentclass[apjl]{emulateapj}

\slugcomment{V5.0, accepted for publications in ApJ (2009), 706, L205}

\shorttitle{Black hole vs.\ stellar mass in COSMOS}
\shortauthors{K.\ Jahnke et al.}

\begin{document}

\title{Massive galaxies in COSMOS: Evolution of black hole vs.\ bulge mass but not
  vs.\ total stellar mass over the last 9 Gyrs?\footnotemark[0]}

\author{Knud Jahnke\altaffilmark{1}}
\author{Angela Bongiorno\altaffilmark{2,3}}
\author{Marcella Brusa\altaffilmark{2}}
\author{Peter Capak\altaffilmark{4}}
\author{Nico Cappelluti\altaffilmark{2}}
\author{Mauricio Cisternas\altaffilmark{1}}
\author{Francesca Civano\altaffilmark{5}}
\author{James Colbert\altaffilmark{4}}
\author{Andrea Comastri\altaffilmark{6}}
\author{Martin Elvis\altaffilmark{5}}
\author{G\"unther Hasinger\altaffilmark{7}}
\author{Olivier Ilbert\altaffilmark{8}}
\author{Chris Impey\altaffilmark{9}}
\author{Katherine Inskip\altaffilmark{1}}
\author{Anton M.\ Koekemoer\altaffilmark{10}}
\author{Simon Lilly\altaffilmark{11}}
\author{Christian Maier\altaffilmark{11}}
\author{Andrea Merloni\altaffilmark{2,12}}
\author{Dominik Riechers\altaffilmark{4,13}}
\author{Mara Salvato\altaffilmark{4,7}}
\author{Eva Schinnerer\altaffilmark{1}}
\author{Nick Z.\ Scoville\altaffilmark{4}}
\author{John Silverman\altaffilmark{11}}
\author{Yoshi Taniguchi\altaffilmark{14}}
\author{Jonathan R.\ Trump\altaffilmark{9}}
\author{Lin Yan\altaffilmark{4}}

\email{jahnke@mpia.de}

\altaffiltext{1}{Max-Planck-Institut f\"ur Astronomie,
K\"onigstuhl 17, D-69117 Heidelberg, Germany}
\altaffiltext{2}{Max-Planck-Institut f\"ur Extraterrestrische Physik,
  Giessenbachstrasse, D-85741 Garching b.\ M\"unchen, Germany}
\altaffiltext{3}{University of Maryland, Baltimore County, 1000 Hilltop
  Circle, Baltimore, MD\,21250, USA}
\altaffiltext{4}{California Institute of Technology, 1200 East California
  Boulevard, MC 249-17, Pasadena, CA\,91125, USA}
\altaffiltext{5}{Harvard Smithsonian Center for Astrophysics, 60 Garden
  St., Cambridge, MA\,02138, USA}
\altaffiltext{6}{INAF-Osservatorio Astronomico di Bologna, via Ranzani 1,
  I-40127 Bologna, Italy}
\altaffiltext{7}{Max-Planck-Institut f\"ur Plasmaphysik,
Boltzmanstrasse 2, D-85741 Garching}
\altaffiltext{8}{Institute for Astronomy, 2680 Woodlawn Dr., University of
  Hawaii, Honolulu, HI\,96822, USA} 
\altaffiltext{9}{Steward Observatory, University of Arizona, 933 North
  Cherry Avenue, Tucson, AZ\,85721}
\altaffiltext{10}{Space Telescope Science Institute,
3700 San Martin Drive, Baltimore, MD\,21218, USA}
\altaffiltext{11}{Department of Physics, ETH Z\"urich, CH-8093 Z\"urich,
  Switzerland} 
\altaffiltext{12}{Excellence Cluster Universe, TUM, Boltzmannstr.\ 2,
  D-85748, Garching, Germany}
\altaffiltext{13}{Hubble Fellow}
\altaffiltext{14}{Research Center for Space and Cosmic Evolution, Ehime
  University, Bunkyo-cho, Matsuyama 790-8577, Japan}

\footnotetext[0]{
Based on observations with the NASA/ESA {\em Hubble Space Telescope},
obtained at the Space Telescope Science Institute, which is operated by
AURA Inc, under NASA contract NAS 5-26555, the XMM-Newton telescope, an
ESA science mission with instruments and contributions directly funded by
ESA Member States and NASA, the European Southern Observatory under Large
Program 175.A-0839, the Magellan Telescope which is operated by the
Carnegie Observatories, and the Subaru Telescope, which is operated by the
National Astronomical Observatory of Japan.
}

\begin{abstract}
We constrain the ratio of black hole (BH) mass to total stellar mass of
type-1 AGN in the COSMOS survey at $1<z<2$. For 10 AGN at mean redshift
$z\sim1.4$ with both HST/ACS and HST/NICMOS imaging data we are able to
compute the total stellar mass $M_\mathrm{*,total}$, based on restframe
UV-to-optical host galaxy colors which constrain mass-to-light ratios. All
objects have virial $M_\mathrm{BH}$-estimates available from the COSMOS
Magellan/IMACS and zCOSMOS surveys. We find within errors zero difference
between the $M_\mathrm{BH}$--$M_\mathrm{*,total}$-relation at $z\sim1.4$
and the $M_\mathrm{BH}$--$M_\mathrm{*,bulge}$-relation in the local
Universe.
Our interpretation is: (a) If our objects were purely bulge-dominated, the
$M_\mathrm{BH}$--$M_\mathrm{*,bulge}$-relation has not evolved since
$z\sim1.4$.
However, (b) since we have evidence for substantial disk components, the
bulges of massive galaxies ($M_\mathrm{*,total}=11.1\pm0.3$ or $\log
M_\mathrm{BH}\sim8.3\pm0.2$) must have grown over the last 9 Gyrs
predominantly by redistribution of disk- into bulge-mass.
Since all necessary stellar mass exists in the galaxy at $z=1.4$, no
star-formation or addition of external stellar material is required, only
a redistribution e.g.\ induced by minor and major merging or through disk
instabilities.
Merging, in addition to redistributing mass in the galaxy, will add both
BH and stellar/bulge mass, but does not change the overall final
$M_\mathrm{BH}$/$M_\mathrm{*,bulge}$ ratio.

Since the overall cosmic stellar and BH mass buildup trace each other
tightly over time, our scenario of bulge-formation in massive galaxies
is {\em independent} of any strong BH-feedback and means that the
mechanism coupling BH and bulge mass until the present is very indirect.
\end{abstract}


\section{Introduction}
Masses of galactic bulges and their central black holes (BHs) follow a
tight relation in the local Universe \citep[e.g.][]{marc03,haer04} with
only 0.3~dex scatter -- strong evidence for a coupled formation and
evolution of galaxies and BHs. The source of this coupling is unclear, but
feedback mechanisms have been proposed involving the central potential
well depth regulating BH accretion, or more violent feedback from active
galactic nuclei (AGN) into their host galaxies
\citep[e.g.][]{hopk06c,some08,menc08}. While these scenarios potentially
provide ingredients for acquiring consensus with observations, all such
models include ad hoc assumptions and do not work from first
principles. Empirical constraints are urgently needed to investigate the
actual physical processes involved in the coupled evolution.

One strong constraint is the evolution of the
$M_\mathrm{BH}$--$M_\mathrm{*,bulge}$-relation over time.  While
circumstantial evidence grows that the value of
$M_\mathrm{BH}$/$M_\mathrm{*,bulge}$ was larger at earlier cosmic times
\citep{peng06a,peng06b,treu07,woo08,walt04,riec08a,riec08b,riec09},
studies are subject to biases \citep{laue07} and better statistics are
required to investigate where in $M_\mathrm{BH}$, or when in cosmic time, a
turnoff from the local $M_\mathrm{BH}$--$M_\mathrm{*,bulge}$-relation
occurs.

Broad-line AGN and their host galaxies are the only systems at higher
redshifts in which both the mass of the galaxy or its bulge as well as its
central BH can be estimated. Here we set constraints on an evolving
$M_\mathrm{BH}$--$M_\mathrm{*,bulge}$-relation by computing optical
color-based stellar masses (from HST/ACS and HST/NICMOS) and combine them
with virial $M_\mathrm{BH}$ (from Magellan/IMACS and zCOSMOS/VLT/VIMOS)
for 10 AGN in the redshift interval $1.06<z<1.92$, 3.2--5.5~Gyrs after the
Big Bang.

Throughout we use AB zero-points unless otherwise noted and a cosmology of
$H_0=70\,\mathrm{km\,s^{-1} Mpc^{-1}}$, $\Omega_M=0.3$, and
$\Omega_\Lambda=0.7$.


\section{Database}
In order to control selection effects on $M_\mathrm{*}$, $M_\mathrm{BH}$
or any special relation between these two, we require a transparent sample
definition.  Our selection of type-1 AGN is based on X-ray detection in
the XMM-COSMOS survey \citep{hasi07,capp09} and subsequent identification
of their optical counterparts \citep{brus07}. Classification as type-1 AGN
for this study uses both spectroscopic identification of
broad emission lines in the Magellan/IMACS \citep{trum09a} and zCOSMOS
\citep{lill07} surveys as well as photometric classification using the
long spectral baseline SED covered in COSMOS \citep[Capak et al., in
  prep.]{capa07,salv09,ilbe09}. As such, only objects with
high-confidence classification enter our sample.

For $\sim$550 type-1 AGN selected this way we require the
following data, resulting in a random subsample: (1) coverage by ACS
F814W \citep{scov07b,koek07}, and (2) NICMOS
parallels\footnote{http://irsa.ipac.caltech.edu/data/COSMOS/images/nicmos/},
as well as (3) spectra for virial $M_\mathrm{BH}$-estimates from the MgII
broad emission line.

The limiting factors are the relative NIC3 coverage of 6.4\% of the ACS
area ($\sim$20 AGN) and the status of ongoing spectroscopic surveys. Black
hole masses have been calculated by \citet{trum09b} from IMACS spectra and
by \citet{merl09} using zCOSMOS/VIMOS. For this letter we have BH masses
available for 10 AGN, spanning the redshift range $1<z<2$. Two AGN were
observed by both instruments -- the $M_\mathrm{BH}$-estimates are
consistent within 0.2~dex in both cases. Sample information is listed in
Table~\ref{tab:all}, including BH masses and galaxy parameters derived
below.

\begin{deluxetable*}{crcccccccc}
\tabletypesize{\scriptsize}
\tablewidth{0pt}
\tablecaption{Sample summary\label{tab:all}}
\tablecolumns{10}
\tablehead{
 \colhead{}&\colhead{} &\colhead{}&\colhead{$M_\mathrm{BH}$\tablenotemark{b}}&\colhead{} &
 \colhead{$F814W_\mathrm{host}$}&\colhead{$F160W_\mathrm{host}$}
 &\colhead{S\'ersic $n$\tablenotemark{d}}&\colhead{$(B-V)_\mathrm{host}^\mathrm{rest}$}& 
 \colhead{$\log(M_\mathrm{*,total})$\tablenotemark{e}}\\
\colhead{XMM-Newton name\tablenotemark{a}}
&\colhead{XID\tablenotemark{a}}&\colhead{$z$}&\colhead{($M_\odot$)}&
\colhead{Ref\tablenotemark{c}}&\colhead{(AB mag)}&\colhead{(AB mag)}&
\colhead{$F160W$}&\colhead{(Vega mag)}&\colhead{($M_\odot$)}
}
\startdata
XMMU J100118.5+022739&  14& 1.065&  8.52& 1   &20.70 &19.41&1.9 &0.38      & 11.18\\
XMMU J100046.8+020016&  59& 1.923&  8.72& 2   &--    &20.28&0.6 &0.15--0.75& 11.07--12.12\\
XMMU J095927.7+020010& 219& 1.248&  8.07& 2   &22.55 &20.61&2.1 &0.49      & 11.00\\
XMMU J100035.3+024303& 281& 1.177&  8.25& 1,2 &22.63 &20.25&3.5 &0.74      & 11.44\\
XMMU J095928.5+015934& 329& 1.166&  8.05& 1,2 &22.95 &21.03&1.7 &0.58      & 10.90\\
XMMU J100130.7+021147&2148& 1.526&  8.43& 1   &23.45 &21.15&1.5 &0.40      & 10.88\\
XMMU J100243.8+020502&2261& 1.260&  8.05& 1   &--    &21.38& -- &0.15--0.75& 10.25--11.09\\
XMMU J100226.9+015938&2637& 1.630&  8.35& 1   &24.12 &20.72&1.8 &0.73      & 11.64\\
XMMU J095903.2+022001&5049& 1.131&  8.40& 2   &22.66 &20.82&1.5 &0.57      & 10.93\\
XMMU J095908.1+024310&5230& 1.359&  8.22& 1   &--    &19.13& -- &0.15--0.75& 11.21--12.08\\
\enddata
\tablenotetext{a}{Original XMM-Newton source name and ID \citep{capp09}}
\tablenotetext{b}{Mean value where two measurements are available}
\tablenotetext{c}{Source for $M_\mathrm{BH}$: (1) Magellan/IMACS
  \citep{trum09b}; (2) zCOSMOS/VIMOS \citep{merl09};
  $M_\mathrm{BH}$-errors are quoted as 0.4~dex and 0.3~dex, respectively}
\tablenotetext{d}{From free-$n$ fit before fixing}
\tablenotetext{e}{Total uncertainty is $\pm$0.4~dex
  (Section~\ref{sec:uncertainty}).}

\end{deluxetable*}


\section{Host galaxy masses}

\subsection{Observed host galaxy photometry and colors}
We obtain information on the host galaxies using broad-band photometry
from the high-resolution HST ACS/WFC images in the F814W (=$I$)
filter with 0\farcs03/pixel sampling \citep{koek07}, and the NICMOS/NIC3
parallels in the F160W (=$H$) filter at 0\farcs101/pixel, both integrated
for one orbit.

In the NICMOS $H$-band the host galaxies of all AGN are clearly resolved,
visible already to the unaided eye. We hence extract the host galaxy flux
from the composite galaxy+AGN NICMOS image by modelling the
two-dimensional light distribution of each object using GALFIT
\citep{peng02,peng09} in Version~3.0 (C.~Y.\ Peng, priv.\ comm.). We
restrict our models to a point-source plus a single elliptical S\'ersic
(\citeyear{sers68}) profile. Previous simulations show that at our
resolution and depth it is unreliable to use many and/or complex galaxy
components \citep{simm08,sanc04a}. We carry out several passes of GALFIT,
first with free S\'ersic parameter\footnote{$n=1$ represents an
  exponential disk, $n=4$ a de Vaucouleurs spheroid.}  $n$, and
subsequently fixed $n = 1$, 2 or 4, depending on the best free-$n$-fit.

We require GALFIT to converge on a sensible solution, indicating that an
actual host galaxy is being described and not e.g.\ uncertainties in the
point spread function (PSF). This means we require a GALFIT
nucleus-to-host contrast of $<3.25$~mag, a half-light radius of $r>2$
pixel, and $0.5<n<8$.

XID~2261 and 5230 both show unrealistically compact host galaxy model
scale-lengths, indicating an unsuccessful GALFIT model. For these two
cases we simply subtract the best fitting single point-source model --
without a S\'ersic component -- from the original image, resulting in an
only slightly oversubtracted/underestimated host galaxy. We use aperture
photometry on the host galaxy in these cases, and the GALFIT galaxy model
magnitude for the other eight.

\begin{figure*}
\begin{center}
\includegraphics[width=15cm,clip]{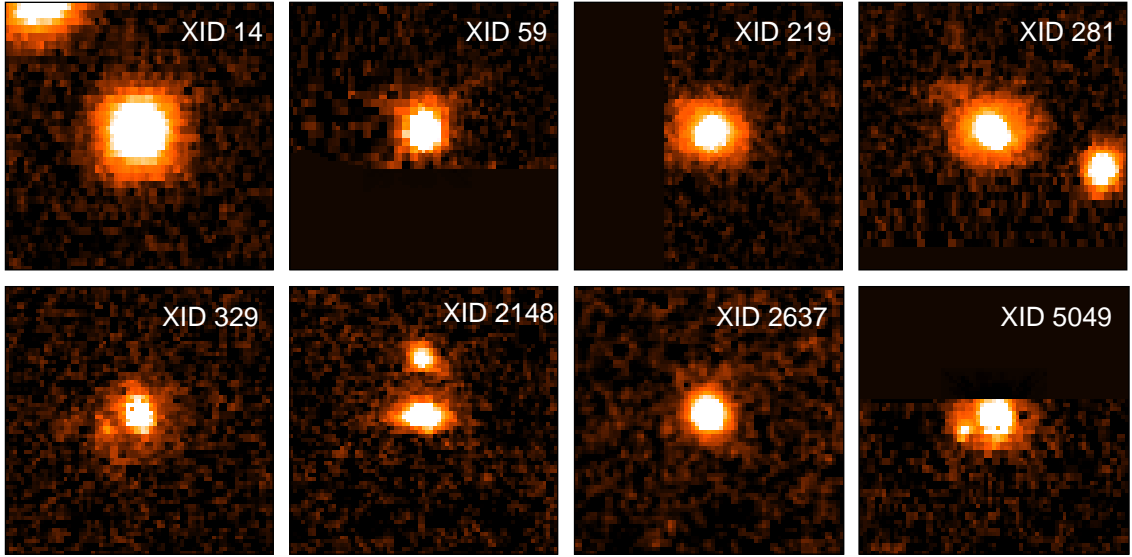}
\end{center}
\caption{\label{fig:allobj} 
Nucleus-removed host galaxies: Galaxy plus nucleus model fitted for 8/10
objects (HST/NIC3 F160W). Images are $7\arcsec\times7\arcsec$, some
objects lie near NICMOS tile edges.
}
\end{figure*}

\begin{figure}
\begin{center}
\includegraphics[width=7.5cm,clip]{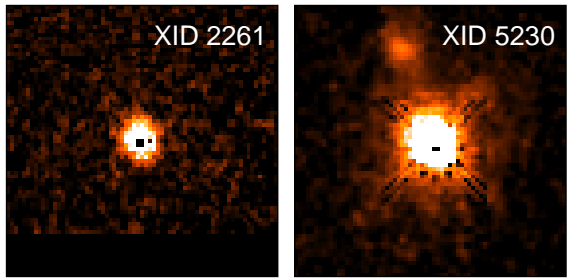}
\end{center}
\caption{\label{fig:allobj2} 
As Figure~\ref{fig:allobj}: XID~2261 and 5230, only nucleus model fitted
and subtracted.
}
\end{figure}

In total we resolve {\em all} 10 host galaxies in the sample -- no object
drops from the sample due to high nucleus-to-host contrast or other
reasons. Extracted host galaxy images are shown in
Figures~\ref{fig:allobj}+\ref{fig:allobj2}.
\medskip

In the ACS $I$-band data the contrast between AGN and host galaxy is less
favorable than in the $H$-band, as expected from the near-UV
SED. Therefore we take a two-step approach. First, we carry out ``peak
subtraction'' removal of the nuclear component for the $I$-band, by
scaling a PSF to the central 4 pixel aperture flux. From statistics on the
expected random residuals after subtraction of this scaled PSF for several
1000 stars \citep[][]{jahn04b}, we require a residual flux of $>$5\% for a
host galaxy to have a high probability of being resolved. This is the case
for 7/10 objects -- the hosts of XIDs~59, 2261, and 5230 remain unresolved
in the ACS image.

The seven resolved objects are again modeled using GALFIT, with S\'ersic
$n$ fixed from the $H$-band fit, thus minimizing S/N-dependent biases in
our extracted colors. As for the $H$-band we check the models for
successful convergence, which is the case for all seven resolved objects,
and again use the host model magnitude. The resulting host galaxy
photometry is also listed in Table~\ref{tab:all}.

\subsection{Host galaxy stellar masses}
Our general approach is to constrain the mass-to-light ratio (M/L) of each
host galaxy by a single optical color. We have successfully employed this
method before to study stellar populations of low-$z$ quasar host
galaxies \citep{jahn04a} or stellar ages and masses of quasar hosts at
$0.5<z<3$ \citep{sanc04a,jahn04b,kuhl05,schr08}. Here we compute stellar
masses from the rest-frame $V$-band luminosity in combination with the M/L
from the rest-frame $(B-V)$ color:

\begin{equation}
  L_\mathrm{V} = 10^{-0.4(V-4.82)}
\end{equation}

\begin{equation} \label{eqn:mass}
  M_\mathrm{*,total} = 10^{-0.952+1.710(B-V)}\times L_\mathrm{V},
\end{equation}

with $L$ and $M$ in solar units. This calibration is based on
template-fitted masses \citep[Chabrier IMF]{bruz03} and luminosities
derived for galaxies in the COSMOS field \citep{ilbe09b}. We convert
luminosities to Vega zeropoint and apply a $-$0.124~dex mass offset to
transform to the mass scale of the models by Charlot \& Bruzual (2007/2009
in prep.) that include contributions from TP-AGB stars. The linear
relation of eqn.~\ref{eqn:mass} is a fit to galaxies in redshift-
($1.0<z<1.6$) and color-range ($0.38<(B-V)<0.74$) of our 7 host galaxies
resolved in both HST bands. The scatter of the fit corresponds to an RMS
uncertainty in resulting stellar mass of $\pm$0.3~dex.

We convert our measured (F814W--F160W) colors to restframe
$(B-V)_\mathrm{host}$ by applying both $K$-correction and a color term,
individually for each object and its redshift. For this purpose we
identify for each galaxy the single stellar population model again from
Charlot \& Bruzual (Chabrier IMF, solar metallicity) with the closest
(F814W--F160W) at a given $z$. Since the interpolation intervals are rather
small and we are not using the interpolation SED to extract any
further information, this method is quite insensitive to the exact choice
of models, and errors from M/L-calibration dominate.

For the three host galaxies only detected in F160W we have to make
assumptions for the interpolation template to convert observed F160W to
$L_\mathrm{V}$, as well as the color (or M/L) in
equation~\ref{eqn:mass}. On the red/old side we assume a value of
$(B-V)=0.75$ (Vega ZP) corresponding to the red end of the red sequence at
$z\sim1.5$ \citep{krie08}, a value also consistent with the reddest values
of the rest of the sample. As a blue limit we use $(B-V)=0.15$,
corresponding to 3$\sigma$ from the mean value for COSMOS inactive
galaxies at $z\sim1.4$, $\log M_\mathrm{*,total}>10.7$ and age
$t<1$~Gyr. These assumptions should bracket the true M/L, and provide
robust limits on the host galaxies' stellar mass.

\subsection{Stellar mass uncertainties}\label{sec:uncertainty}
Uncertainties in stellar mass have several sources. In addition to the
$(B-V)$ calibration stated above, the strongest contributions come from
(1) GALFIT precision of extracting the host galaxy, (2) potential
influence from different spatial resolutions in the ACS and NIC3 images,
and (3) dependency of bandpass conversions on assumed templates.

(1) \citet{sanc04a} tested how reliably GALFIT can derive host
galaxy photometry in one-orbit ACS data. Given the typical host galaxy
magnitude of our sample in the F814W filter we conclude an uncertainty of
0.15~mag for F814W, and -- due to counteracting effects of lower
spatial resolution but more favorable galaxy-to-nucleus contrast -- also for
F160W.

(2) GALFIT's functionallity depends on the spatial resolution of an image
and the spatial difference between the AGN and galaxy component. If a
galaxy is compact and the PSF not well characterized, flux transfer
between components is possible, its amplitude potentially depending on
spatial resolution.
We test if the different resolutions of the ACS and NIC3 have a
significant influence, by rebinning a mock AGN+galaxy image resembling a
typical object to different spatial resolutions. The recovered host galaxy
photometry shows an rms variation of $\sim$0.15~mag, no systematic offset,
and a negligible trend with resolution. This rms scatter is consistent
with the GALFIT precision from (1) above and we conclude that resolution
effects are insignificant.

(3) The bandpass conversion and $K$-correction depend on the IMF and
metallicity of the assumed single population interpolation model. The
masses derived from Chabrier and Salpeter IMF generally differ by less than
0.1~dex. The metallicity of galaxies of the estimated masses at $1<z<2$ is
expected to range between solar and 3$\times$ solar
\citep[$8.8<12+\log(O/H)<9.2$,][]{trem04,maio08}. Changing from the solar
metallicity used, to the most metal-rich templates ($Z=0.05$), masses
change by $\la$0.05~dex. We conclude that our stellar masses should be
good to within $\sim$0.1~dex from the choice of bandpass conversion SEDs.
\medskip

In combination we find our error budget in stellar mass is 0.21~mag for
the $(B-V)$-color (0.15~mag from each band), corresponding to 0.27~dex
uncertainty in stellar mass -- dominating the three sources of uncertainty
above. Adding the uncertainty of the mass calibration from
equation~\ref{eqn:mass}, the total uncertainty in stellar mass is
$\pm$0.4~dex.


\section{Results}

\begin{figure}
\centerline{\includegraphics[height=8cm,clip]{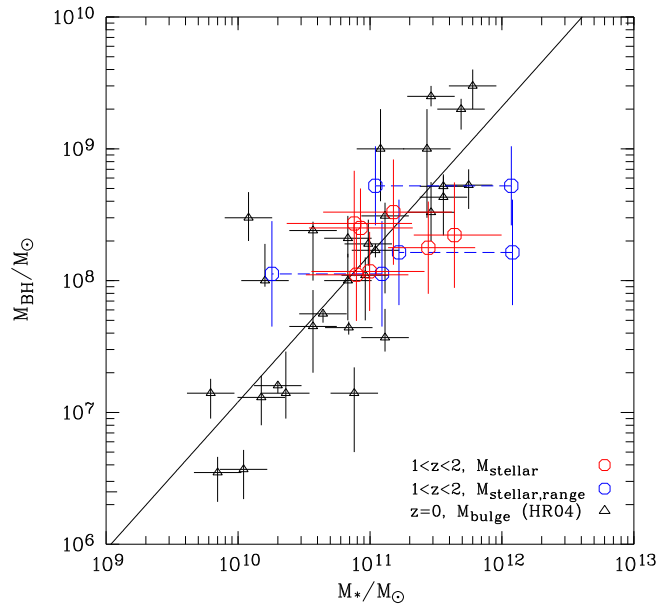}}
\caption{\label{fig:mbh_mstell} 
$M_\mathrm{BH}$--$M_\mathrm{*,total}$-relation from COSMOS ACS+NIMOS:
  Shown are ten type-1 AGN in the redshift range $1.06<z<1.92$, seven with
  direct stellar mass estimates (red circles) and three with bracketing
  range (blue circles and lines). The points are overplotted over the
  local $M_\mathrm{BH}$--$M_\mathrm{*,bulge}$-relation by \citet{haer04}
  (black triangles) and its best fit (solid line, $\log M_\mathrm{BH} =
  8.2 + 1.12\times(\log M_\mathrm{*,bulge}-11)$). At a given mass there is
  no difference in the $M_\mathrm{BH}$/$M_\mathrm{*,total}$-ratio at
  $z\sim1.4$ and local $M_\mathrm{BH}$/$M_\mathrm{*,bulge}$-ratio, for the
  sampled BH mass range, $\log M_\mathrm{BH}\sim8.3$.
}
\end{figure}

While we can not estimate bulge masses directly, we find that the seven
objects with direct {\em total stellar mass} estimates fall directly onto
the $M_\mathrm{BH}$--$M_\mathrm{*,bulge}$-relation
(Fig.~\ref{fig:mbh_mstell}) of the local universe from \citet{haer04}. The
three objects with a bracketing range on stellar mass are also consistent
with the local relation. The objects have a maximum deviation of
$\sim$0.3~dex perpendicular to the $z=0$ relation. The seven non-limit AGN
in the $z$-range $1.06<z<1.65$ show a mean ratio
$M_\mathrm{BH}/M_\mathrm{*,total}=0.00178\pm0.0012$ and mean
$\log(M_\mathrm{BH})=8.31$. This is consistent with the value at $z=0$ of
$M_\mathrm{BH}/M_\mathrm{*,bulge}=0.00165$ at the same $M_\mathrm{BH}$,
and has exactly the same 0.3dex scatter.

\citet{merl09} compute stellar masses for a larger sample of COSMOS type-1
AGN, using an independent SED-decomposition method. For 18 galaxies where
stellar masses could be estimated with both methods (five objects are part
of this study, 13 have no BH mass estimates yet) their masses are smaller
by 0.1--0.2~dex. This agreement within the error bars reinforces our
conclusion that our mass estimates are robust. In total they find a mild
deviation from the $z=0$ relation, but most of their signal comes from
objects at $z>1.5$, which is not well covered by our study.

\section{Discussion \& Conclusions}
\subsection{Completeness and systematics}
Are there systematic effects inherent in the dataset that prevent us from
detecting objects which deviate more strongly from this relation? X-ray
selection and subsequent broad-line AGN classification finds all AGN down
to a contrast of AGN/galaxy$\la$10\%, beyond which the stellar light
swamps any AGN signature. The general limit is set by the magnitude limit
of the spectral followup, modified by this contrast. At the observed
$M_\mathrm{BH}$ we expect generally high accretion rates
\citep[e.g.][]{zhen09}, thus this detection limit converts to a limit in
$M_\mathrm{BH}$ present in the data. The additional ACS and NIC3 coverage
are random processes. Being able to extract a host galaxy for every single
object of our sample means that aside from the $M_\mathrm{BH}$-limit we
have no completeness limitations with respect to galaxy mass. In summary,
we do not expect missing objects in the $M_\mathrm{BH}$-regime currently
populated.
\medskip

We also comment on the luminosity function or ``Lauer'' bias
\citep{laue07}: The $M_\mathrm{BH}$--$M_\mathrm{*,total}$-relation has a
scatter and the BH luminosity function drops rapidly towards higher
$L$. In combination, a flux-limited sample of AGN will select many more
massive BHs in lower mass galaxies than vice versa, biasing the sample
towards higher $M_\mathrm{BH}$.

Despite the local comparison sample not being selected by $M_\mathrm{BH}$
-- it consists of inactive galaxies -- we do not expect a large bias: The
\citet{merl09} and \citet{trum09b} BH masses follow the mass calibration
by \citet{onke04}. Their virial $M_\mathrm{BH}$ for active galaxies are
calibrated by a forced match to the $M_\mathrm{BH}$--$\sigma_*$-relation for
inactive galaxies. This might create ``wrong'' $M_\mathrm{BH}$ estimates,
but it compensates most of the expected offset $\Delta M_\mathrm{BH}$
between our AGN-sample and any (active or inactive) sample in the local
universe.  Together with only a small scatter in the local
$M_\mathrm{BH}$--$M_\mathrm{*,bulge}$ it is not surprising to find no
discernable signature of the LF-bias in our data.

\subsection{For Massive Galaxies: Relative Non-Evolution of BH Mass Versus Total Stellar Mass}

\citet{haer04} derive dynamical {\em bulge} masses -- we derive {\em
  total} stellar masses of the host galaxies. Interpreting the coincidence
of these two relations at $z=0$ and $z\sim1.4$ depends on how much stellar
mass in our sample is actually part of a disk component. The stellar
masses of $\log M_\mathrm{*,total}=11.1\pm0.3$ are already high in the
galaxy mass function and our sample galaxies are the likely progenitors of
giant ellipticals in the local universe. If they were already
bulge-dominated at the observed redshifts, nine billion years earlier, our
total masses revert to bulge masses and our observations mean zero
evolution (though consistent with evolution of factor $<$2.65 within
errors) in $M_\mathrm{BH}$--$M_\mathrm{*,bulge}$, at these
$M_\mathrm{BH}$.

\subsection{Evolution of BH Mass Versus Bulge Mass and Velocity Dispersion}
While the actual bulge-to-total mass-ratio $(B/T)_\mathrm{mass}$ of our
galaxies -- in fact of any galaxy at these redshifts -- is unknown, we
have circumstantial evidence that our galaxies could contain substantial
disk components: (a) Visual impression in
Figs.~\ref{fig:allobj}+\ref{fig:allobj2}, (b) some galaxies have S\'ersic
indices near $n=1$ (Table~\ref{tab:all}), (c) some best fitting SSP models
have ages down to 1~Gyr (oldest: 5~Gyr), (d) the $z=0.36$ host galaxies of
\citet{treu07}, with similar $M_\mathrm{BH}$, have
$(B/T)_\mathrm{mass}\sim1/3$, (e) more than 50\% of $\log
M_\mathrm{*,total}=11.1\pm0.3$ galaxies are strongly starforming at
$z\sim1$ \citep{noes07} and (f) of them $\la40\%$ are on the red sequence
\citep[at $z=1.5$,][]{tayl09}.

With a substantial disk component a different interpretation is possible:

\begin{itemize}
\item We know, independently of their exact evolutionary path, these
  galaxies have to end up on the local
  $M_\mathrm{BH}$--$M_\mathrm{*,bulge}$-relation.
 If they currently obey the same relation but with
  $M_\mathrm{*,total}$, then their masses in bulge+disk at $z=1.4$ can end
  up in their bulges 9~Gyrs later with no addition of mass to either the
  BH or the bulge from outside the galaxy or from star-formation.

\item What is required is a process redistributing disk stars to the
  bulge: major and minor galaxy merging \citep{hopk09} and disk
  instabilities \citep{parr09} together dominate bulge-creation at high
  masses. 

\item In addition to the disk-to-bulge conversion process these galaxies can
  still grow. Merging of similar systems will just coadd BHs and stellar
  components separately, moving the system parallel along the local
  scaling relation. However total mass growth is limited since the
  observed evolution in space density of massive galaxies at $z<1.5$ is
  small \citep[e.g.,][and references therein]{ilbe09}.

\item Even wet mergers are allowed: The gas conversion efficiency is not
  very high -- in individual mergers \citep{crot06b} and also overall
  \citep[at $z\sim0.6$,][]{roba09} only 10\% of SF arise from merging.

\item Any subsequent (or ongoing) AGN phase will also less than double the
  BH mass, since high accretion states are rare and short
  \citep{hopk06c}. Any BH or stellar mass change below a factor of two
  will lie within the 0.3~dex scatter of the local relation.

\end{itemize}

In the case of a disk component in our galaxies at $z=1.4$, bulge masses are
smaller than the total stellar mass in Fig.~\ref{fig:mbh_mstell} and the
$M_\mathrm{BH}$--$M_\mathrm{*,bulge}$-relation will actually evolve, when
the galaxy-structures are changing over time through merging. They will
constantly move closer to the local relation -- consistent with
predictions from simulations of a merger-driven bulge evolution
\citep{crot06b}.

This has the implication that our result is consistent with the
non-evolution of the {\em bulge} mass relation at $z<1.7$ for massive
early type quasar host galaxies found by \citet{peng06a,peng06b} -- there
$M_\mathrm{*,bulge}=M_\mathrm{*,total}$. At the same time we agree with
the strong evolution in $M_\mathrm{BH}$--$M_\mathrm{*,bulge}$ and
$M_\mathrm{BH}$--$\sigma_\mathrm{*,bulge}$ claimed by \citet{treu07} and
\citet{woo08}: If we assume as a limit for our galaxies
$(B/T)_\mathrm{mass}\le1/3$, our results predict an evolution of
$M_\mathrm{BH}$/$M_\mathrm{*,bulge}\le (1+z)^{1.2}$, consistent with
$M_\mathrm{BH}$/$M_\mathrm{*,bulge}\le (1+z)^{1.5\pm1.0}$ found by
\citet{treu07} for Seyfert 1 galaxies of the same black hole mass at
$z=0.36$.

This path of converging co-evolution of bulge and BH is fully indepedent
of any interaction or feedback {\em between} bulge and BH -- the AGN can
even be switched off since $z=1.4$. What we are witnessing is the formation of
the bulge independent of the BH. Only before $z=1.4$ is a mechanism
required to connect total stellar mass and BH mass in massive galaxies. We
might see an indication of this in the large offset of the relation for
$z>4$ \citep{walt04,riec08a,riec08b,riec09}, but the Lauer-bias and the
different mass scale ($M_\mathrm{BH}>9.2$) complicate the picture. Since
stellar mass and BH mass buildup trace each other very well over cosmic
time with a factor similar to the local ratio of BH and stellar mass
\citep{zhen09}, the coupling mechanism can be very indirect and does not
need to be dominated by a strong version of AGN feedback.


\acknowledgments
KJ would like to thank E.~F.\ Bell, N.\ Neumayer, C.~Y.\ Peng, and
A.~van~der~Wel for very fruitful discussions.
KJ is supported through the Emmy Noether Programme of the German Science
Foundation (DFG).
DR acknowledges support from from NASA through a Hubble Fellowship.

{\it Facilities:} \facility{ESO VLT (VIMOS)}, \facility{HST (ACS, NICMOS)},
\facility{Magellan (IMACS)}, \facility{XMM-Newton}.


\begin{thebibliography}{48}
\expandafter\ifx\csname natexlab\endcsname\relax\def\natexlab#1{#1}\fi

\bibitem[{{Brusa} {et~al.}(2007){Brusa}, {Zamorani}, {Comastri}, {Hasinger},
  {Cappelluti}, {Civano}, {Finoguenov}, {Mainieri}, {Salvato}, {Vignali},
  {Elvis}, {Fiore}, {Gilli}, {Impey}, {Lilly}, {Mignoli}, {Silverman}, {Trump},
  {Urry}, {Bender}, {Capak}, {Huchra}, {Kneib}, {Koekemoer}, {Leauthaud},
  {Lehmann}, {Massey}, {Matute}, {McCarthy}, {McCracken}, {Rhodes}, {Scoville},
  {Taniguchi}, \& {Thompson}}]{brus07}
{Brusa}, M., {Zamorani}, G., {Comastri}, A., {Hasinger}, G., {Cappelluti}, N.,
  {Civano}, F., {Finoguenov}, A., {Mainieri}, V., {Salvato}, M., {Vignali}, C.,
  {Elvis}, M., {Fiore}, F., {Gilli}, R., {Impey}, C.~D., {Lilly}, S.~J.,
  {Mignoli}, M., {Silverman}, J., {Trump}, J., {Urry}, C.~M., {Bender}, R.,
  {Capak}, P., {Huchra}, J.~P., {Kneib}, J.~P., {Koekemoer}, A., {Leauthaud},
  A., {Lehmann}, I., {Massey}, R., {Matute}, I., {McCarthy}, P.~J.,
  {McCracken}, H.~J., {Rhodes}, J., {Scoville}, N.~Z., {Taniguchi}, Y., \&
  {Thompson}, D. 2007, \apjs, 172, 353

\bibitem[{{Bruzual} \& {Charlot}(2003)}]{bruz03}
{Bruzual}, G., \& {Charlot}, S. 2003, \mnras, 344, 1000

\bibitem[{{Capak} {et~al.}(2007){Capak}, {Aussel}, {Ajiki}, {McCracken},
  {Mobasher}, {Scoville}, {Shopbell}, {Taniguchi}, {Thompson}, {Tribiano},
  {Sasaki}, {Blain}, {Brusa}, {Carilli}, {Comastri}, {Carollo}, {Cassata},
  {Colbert}, {Ellis}, {Elvis}, {Giavalisco}, {Green}, {Guzzo}, {Hasinger},
  {Ilbert}, {Impey}, {Jahnke}, {Kartaltepe}, {Kneib}, {Koda}, {Koekemoer},
  {Komiyama}, {Leauthaud}, {Lefevre}, {Lilly}, {Liu}, {Massey}, {Miyazaki},
  {Murayama}, {Nagao}, {Peacock}, {Pickles}, {Porciani}, {Renzini}, {Rhodes},
  {Rich}, {Salvato}, {Sanders}, {Scarlata}, {Schiminovich}, {Schinnerer},
  {Scodeggio}, {Sheth}, {Shioya}, {Tasca}, {Taylor}, {Yan}, \&
  {Zamorani}}]{capa07}
{Capak}, P., {Aussel}, H., {Ajiki}, M., {McCracken}, H.~J., {Mobasher}, B.,
  {Scoville}, N., {Shopbell}, P., {Taniguchi}, Y., {Thompson}, D., {Tribiano},
  S., {Sasaki}, S., {Blain}, A.~W., {Brusa}, M., {Carilli}, C., {Comastri}, A.,
  {Carollo}, C.~M., {Cassata}, P., {Colbert}, J., {Ellis}, R.~S., {Elvis}, M.,
  {Giavalisco}, M., {Green}, W., {Guzzo}, L., {Hasinger}, G., {Ilbert}, O.,
  {Impey}, C., {Jahnke}, K., {Kartaltepe}, J., {Kneib}, J.-P., {Koda}, J.,
  {Koekemoer}, A., {Komiyama}, Y., {Leauthaud}, A., {Lefevre}, O., {Lilly}, S.,
  {Liu}, C., {Massey}, R., {Miyazaki}, S., {Murayama}, T., {Nagao}, T.,
  {Peacock}, J.~A., {Pickles}, A., {Porciani}, C., {Renzini}, A., {Rhodes}, J.,
  {Rich}, M., {Salvato}, M., {Sanders}, D.~B., {Scarlata}, C., {Schiminovich},
  D., {Schinnerer}, E., {Scodeggio}, M., {Sheth}, K., {Shioya}, Y., {Tasca},
  L.~A.~M., {Taylor}, J.~E., {Yan}, L., \& {Zamorani}, G. 2007, \apjs, 172, 99

\bibitem[{{Cappelluti} {et~al.}(2009){Cappelluti}, {Brusa}, {Hasinger},
  {Comastri}, {Zamorani}, {Finoguenov}, {Gilli}, {Puccetti}, {Miyaji},
  {Salvato}, {Vignali}, {Aldcroft}, {B{\"o}hringer}, {Brunner}, {Civano},
  {Elvis}, {Fiore}, {Fruscione}, {Griffiths}, {Guzzo}, {Iovino}, {Koekemoer},
  {Mainieri}, {Scoville}, {Shopbell}, {Silverman}, \& {Urry}}]{capp09}
{Cappelluti}, N., {Brusa}, M., {Hasinger}, G., {Comastri}, A., {Zamorani}, G.,
  {Finoguenov}, A., {Gilli}, R., {Puccetti}, S., {Miyaji}, T., {Salvato}, M.,
  {Vignali}, C., {Aldcroft}, T., {B{\"o}hringer}, H., {Brunner}, H., {Civano},
  F., {Elvis}, M., {Fiore}, F., {Fruscione}, A., {Griffiths}, R.~E., {Guzzo},
  L., {Iovino}, A., {Koekemoer}, A.~M., {Mainieri}, V., {Scoville}, N.~Z.,
  {Shopbell}, P., {Silverman}, J., \& {Urry}, C.~M. 2009, \aap, 497, 635

\bibitem[{{Croton}(2006)}]{crot06b}
{Croton}, D.~J. 2006, \mnras, 369, 1808

\bibitem[{{H{\"a}ring} \& {Rix}(2004)}]{haer04}
{H{\"a}ring}, N., \& {Rix}, H.-W. 2004, \apj, 604, L89

\bibitem[{{Hasinger} {et~al.}(2007){Hasinger}, {Cappelluti}, {Brunner},
  {Brusa}, {Comastri}, {Elvis}, {Finoguenov}, {Fiore}, {Franceschini}, {Gilli},
  {Griffiths}, {Lehmann}, {Mainieri}, {Matt}, {Matute}, {Miyaji}, {Molendi},
  {Paltani}, {Sanders}, {Scoville}, {Tresse}, {Urry}, {Vettolani}, \&
  {Zamorani}}]{hasi07}
{Hasinger}, G., {Cappelluti}, N., {Brunner}, H., {Brusa}, M., {Comastri}, A.,
  {Elvis}, M., {Finoguenov}, A., {Fiore}, F., {Franceschini}, A., {Gilli}, R.,
  {Griffiths}, R.~E., {Lehmann}, I., {Mainieri}, V., {Matt}, G., {Matute}, I.,
  {Miyaji}, T., {Molendi}, S., {Paltani}, S., {Sanders}, D.~B., {Scoville}, N.,
  {Tresse}, L., {Urry}, C.~M., {Vettolani}, P., \& {Zamorani}, G. 2007, \apjs,
  172, 29

\bibitem[{{Hopkins} {et~al.}(2009){Hopkins}, {Bundy}, {Croton}, {Hernquist},
  {Keres}, {Khochfar}, {Stewart}, {Wetzel}, \& {Younger}}]{hopk09}
{Hopkins}, P.~F., {Bundy}, K., {Croton}, D., {Hernquist}, L., {Keres}, D.,
  {Khochfar}, S., {Stewart}, K., {Wetzel}, A., \& {Younger}, J.~D. 2009,
  submitted to MNRAS, arxiv/0906.5357

\bibitem[{{Hopkins} {et~al.}(2006){Hopkins}, {Hernquist}, {Cox}, {Di Matteo},
  {Robertson}, \& {Springel}}]{hopk06c}
{Hopkins}, P.~F., {Hernquist}, L., {Cox}, T.~J., {Di Matteo}, T., {Robertson},
  B., \& {Springel}, V. 2006, \apjs, 163, 1

\bibitem[{{Ilbert} {et~al.}(2009{\natexlab{a}}){Ilbert}, {Capak}, {Salvato},
  {Aussel}, {McCracken}, {Sanders}, {Scoville}, {Kartaltepe}, {Arnouts},
  {Floc'h}, {Mobasher}, {Taniguchi}, {Lamareille}, {Leauthaud}, {Sasaki},
  {Thompson}, {Zamojski}, {Zamorani}, {Bardelli}, {Bolzonella}, {Bongiorno},
  {Brusa}, {Caputi}, {Carollo}, {Contini}, {Cook}, {Coppa}, {Cucciati}, {de la
  Torre}, {de Ravel}, {Franzetti}, {Garilli}, {Hasinger}, {Iovino}, {Kampczyk},
  {Kneib}, {Knobel}, {Kovac}, {LeBorgne}, {LeBrun}, {F{\`e}vre}, {Lilly},
  {Looper}, {Maier}, {Mainieri}, {Mellier}, {Mignoli}, {Murayama}, {Pell{\`o}},
  {Peng}, {P{\'e}rez-Montero}, {Renzini}, {Ricciardelli}, {Schiminovich},
  {Scodeggio}, {Shioya}, {Silverman}, {Surace}, {Tanaka}, {Tasca}, {Tresse},
  {Vergani}, \& {Zucca}}]{ilbe09}
{Ilbert}, O., {Capak}, P., {Salvato}, M., {Aussel}, H., {McCracken}, H.~J.,
  {Sanders}, D.~B., {Scoville}, N., {Kartaltepe}, J., {Arnouts}, S., {Floc'h},
  E.~L., {Mobasher}, B., {Taniguchi}, Y., {Lamareille}, F., {Leauthaud}, A.,
  {Sasaki}, S., {Thompson}, D., {Zamojski}, M., {Zamorani}, G., {Bardelli}, S.,
  {Bolzonella}, M., {Bongiorno}, A., {Brusa}, M., {Caputi}, K.~I., {Carollo},
  C.~M., {Contini}, T., {Cook}, R., {Coppa}, G., {Cucciati}, O., {de la Torre},
  S., {de Ravel}, L., {Franzetti}, P., {Garilli}, B., {Hasinger}, G., {Iovino},
  A., {Kampczyk}, P., {Kneib}, J.-P., {Knobel}, C., {Kovac}, K., {LeBorgne},
  J.~F., {LeBrun}, V., {F{\`e}vre}, O.~L., {Lilly}, S., {Looper}, D., {Maier},
  C., {Mainieri}, V., {Mellier}, Y., {Mignoli}, M., {Murayama}, T.,
  {Pell{\`o}}, R., {Peng}, Y., {P{\'e}rez-Montero}, E., {Renzini}, A.,
  {Ricciardelli}, E., {Schiminovich}, D., {Scodeggio}, M., {Shioya}, Y.,
  {Silverman}, J., {Surace}, J., {Tanaka}, M., {Tasca}, L., {Tresse}, L.,
  {Vergani}, D., \& {Zucca}, E. 2009{\natexlab{a}}, \apj, 690, 1236

\bibitem[{{Ilbert} {et~al.}(2009{\natexlab{b}}){Ilbert}, {Salvato}, {Le
  Floc'h}, {Aussel}, {Capak}, {McCracken}, {Mobasher}, {Kartaltepe},
  {Scoville}, {Sanders}, {Arnouts}, {Bundy}, {Cassata}, {Kneib}, {Koekemoer},
  {Le Fevre}, {Lilly}, {Surace}, {Taniguchi}, {Tasca}, {Thompson}, {Tresse},
  {Zamojski}, {Zamorani}, \& {Zucca}}]{ilbe09b}
{Ilbert}, O., {Salvato}, M., {Le Floc'h}, E., {Aussel}, H., {Capak}, P.,
  {McCracken}, H.~J., {Mobasher}, B., {Kartaltepe}, J., {Scoville}, N.,
  {Sanders}, D.~B., {Arnouts}, S., {Bundy}, K., {Cassata}, P., {Kneib}, J.~.,
  {Koekemoer}, A., {Le Fevre}, O., {Lilly}, S., {Surace}, J., {Taniguchi}, Y.,
  {Tasca}, L., {Thompson}, D., {Tresse}, L., {Zamojski}, M., {Zamorani}, G., \&
  {Zucca}, E. 2009{\natexlab{b}}, submitted to ApJ, arXiv 0903.0102

\bibitem[{Jahnke {et~al.}(2004{\natexlab{a}})Jahnke, Kuhlbrodt, \&
  Wisotzki}]{jahn04a}
Jahnke, K., Kuhlbrodt, B., \& Wisotzki, L. 2004{\natexlab{a}}, MNRAS, 352, 399

\bibitem[{Jahnke {et~al.}(2004{\natexlab{b}})Jahnke, S\'anchez, Wisotzki,
  Barden, Beckwith, Bell, Borch, Caldwell, H\"au{\ss}ler, Heymans, Jogee,
  McIntosh, Meisenheimer, Peng, Rix, Somerville, \& Wolf}]{jahn04b}
Jahnke, K., S\'anchez, S.~F., Wisotzki, L., Barden, M., Beckwith, S.~V.~W.,
  Bell, E.~F., Borch, A., Caldwell, J.~A.~R., H\"au{\ss}ler, B., Heymans, C.,
  Jogee, S., McIntosh, D.~H., Meisenheimer, K., Peng, C.~Y., Rix, H.-W.,
  Somerville, R.~S., \& Wolf, C. 2004{\natexlab{b}}, ApJ, 614, 568

\bibitem[{{Koekemoer} {et~al.}(2007){Koekemoer}, {Aussel}, {Calzetti}, {Capak},
  {Giavalisco}, {Kneib}, {Leauthaud}, {Le F{\`e}vre}, {McCracken}, {Massey},
  {Mobasher}, {Rhodes}, {Scoville}, \& {Shopbell}}]{koek07}
{Koekemoer}, A.~M., {Aussel}, H., {Calzetti}, D., {Capak}, P., {Giavalisco},
  M., {Kneib}, J.-P., {Leauthaud}, A., {Le F{\`e}vre}, O., {McCracken}, H.~J.,
  {Massey}, R., {Mobasher}, B., {Rhodes}, J., {Scoville}, N., \& {Shopbell},
  P.~L. 2007, \apjs, 172, 196

\bibitem[{{Kriek} {et~al.}(2008){Kriek}, {van der Wel}, {van Dokkum}, {Franx},
  \& {Illingworth}}]{krie08}
{Kriek}, M., {van der Wel}, A., {van Dokkum}, P.~G., {Franx}, M., \&
  {Illingworth}, G.~D. 2008, \apj, 682, 896

\bibitem[{Kuhlbrodt {et~al.}(2005)Kuhlbrodt, \"Orndahl, Wisotzki, \&
  Jahnke}]{kuhl05}
Kuhlbrodt, B., \"Orndahl, E., Wisotzki, L., \& Jahnke, K. 2005, \aap, 439, 497

\bibitem[{{Lauer} {et~al.}(2007){Lauer}, {Tremaine}, {Richstone}, \&
  {Faber}}]{laue07}
{Lauer}, T.~R., {Tremaine}, S., {Richstone}, D., \& {Faber}, S.~M. 2007, \apj,
  670, 249

\bibitem[{{Lilly} {et~al.}(2007){Lilly}, {Le F{\`e}vre}, {Renzini}, {Zamorani},
  {Scodeggio}, {Contini}, {Carollo}, {Hasinger}, {Kneib}, {Iovino}, {Le Brun},
  {Maier}, {Mainieri}, {Mignoli}, {Silverman}, {Tasca}, {Bolzonella},
  {Bongiorno}, {Bottini}, {Capak}, {Caputi}, {Cimatti}, {Cucciati}, {Daddi},
  {Feldmann}, {Franzetti}, {Garilli}, {Guzzo}, {Ilbert}, {Kampczyk}, {Kovac},
  {Lamareille}, {Leauthaud}, {Borgne}, {McCracken}, {Marinoni}, {Pello},
  {Ricciardelli}, {Scarlata}, {Vergani}, {Sanders}, {Schinnerer}, {Scoville},
  {Taniguchi}, {Arnouts}, {Aussel}, {Bardelli}, {Brusa}, {Cappi}, {Ciliegi},
  {Finoguenov}, {Foucaud}, {Franceschini}, {Halliday}, {Impey}, {Knobel},
  {Koekemoer}, {Kurk}, {Maccagni}, {Maddox}, {Marano}, {Marconi}, {Meneux},
  {Mobasher}, {Moreau}, {Peacock}, {Porciani}, {Pozzetti}, {Scaramella},
  {Schiminovich}, {Shopbell}, {Smail}, {Thompson}, {Tresse}, {Vettolani},
  {Zanichelli}, \& {Zucca}}]{lill07}
{Lilly}, S.~J., {Le F{\`e}vre}, O., {Renzini}, A., {Zamorani}, G., {Scodeggio},
  M., {Contini}, T., {Carollo}, C.~M., {Hasinger}, G., {Kneib}, J.-P.,
  {Iovino}, A., {Le Brun}, V., {Maier}, C., {Mainieri}, V., {Mignoli}, M.,
  {Silverman}, J., {Tasca}, L.~A.~M., {Bolzonella}, M., {Bongiorno}, A.,
  {Bottini}, D., {Capak}, P., {Caputi}, K., {Cimatti}, A., {Cucciati}, O.,
  {Daddi}, E., {Feldmann}, R., {Franzetti}, P., {Garilli}, B., {Guzzo}, L.,
  {Ilbert}, O., {Kampczyk}, P., {Kovac}, K., {Lamareille}, F., {Leauthaud}, A.,
  {Borgne}, J.-F.~L., {McCracken}, H.~J., {Marinoni}, C., {Pello}, R.,
  {Ricciardelli}, E., {Scarlata}, C., {Vergani}, D., {Sanders}, D.~B.,
  {Schinnerer}, E., {Scoville}, N., {Taniguchi}, Y., {Arnouts}, S., {Aussel},
  H., {Bardelli}, S., {Brusa}, M., {Cappi}, A., {Ciliegi}, P., {Finoguenov},
  A., {Foucaud}, S., {Franceschini}, R., {Halliday}, C., {Impey}, C., {Knobel},
  C., {Koekemoer}, A., {Kurk}, J., {Maccagni}, D., {Maddox}, S., {Marano}, B.,
  {Marconi}, G., {Meneux}, B., {Mobasher}, B., {Moreau}, C., {Peacock}, J.~A.,
  {Porciani}, C., {Pozzetti}, L., {Scaramella}, R., {Schiminovich}, D.,
  {Shopbell}, P., {Smail}, I., {Thompson}, D., {Tresse}, L., {Vettolani}, G.,
  {Zanichelli}, A., \& {Zucca}, E. 2007, \apjs, 172, 70

\bibitem[{{Maiolino} {et~al.}(2008){Maiolino}, {Nagao}, {Grazian}, {Cocchia},
  {Marconi}, {Mannucci}, {Cimatti}, {Pipino}, {Ballero}, {Calura}, {Chiappini},
  {Fontana}, {Granato}, {Matteucci}, {Pastorini}, {Pentericci}, {Risaliti},
  {Salvati}, \& {Silva}}]{maio08}
{Maiolino}, R., {Nagao}, T., {Grazian}, A., {Cocchia}, F., {Marconi}, A.,
  {Mannucci}, F., {Cimatti}, A., {Pipino}, A., {Ballero}, S., {Calura}, F.,
  {Chiappini}, C., {Fontana}, A., {Granato}, G.~L., {Matteucci}, F.,
  {Pastorini}, G., {Pentericci}, L., {Risaliti}, G., {Salvati}, M., \& {Silva},
  L. 2008, \aap, 488, 463

\bibitem[{{Marconi} \& {Hunt}(2003)}]{marc03}
{Marconi}, A., \& {Hunt}, L.~K. 2003, \apjl, 589, L21

\bibitem[{{Menci} {et~al.}(2008){Menci}, {Fiore}, {Puccetti}, \&
  {Cavaliere}}]{menc08}
{Menci}, N., {Fiore}, F., {Puccetti}, S., \& {Cavaliere}, A. 2008, \apj, 686,
  219

\bibitem[{Merloni {et~al.}(2009)Merloni, Bongiorno, {et~al.}}]{merl09}
Merloni, A., Bongiorno, A., {et~al.} 2009, submitted to ApJ

\bibitem[{{Noeske} {et~al.}(2007){Noeske}, {Faber}, {Weiner}, {Koo}, {Primack},
  {Dekel}, {Papovich}, {Conselice}, {Le Floc'h}, {Rieke}, {Coil}, {Lotz},
  {Somerville}, \& {Bundy}}]{noes07}
{Noeske}, K.~G., {Faber}, S.~M., {Weiner}, B.~J., {Koo}, D.~C., {Primack},
  J.~R., {Dekel}, A., {Papovich}, C., {Conselice}, C.~J., {Le Floc'h}, E.,
  {Rieke}, G.~H., {Coil}, A.~L., {Lotz}, J.~M., {Somerville}, R.~S., \&
  {Bundy}, K. 2007, \apjl, 660, L47

\bibitem[{{Onken} {et~al.}(2004){Onken}, {Ferrarese}, {Merritt}, {Peterson},
  {Pogge}, {Vestergaard}, \& {Wandel}}]{onke04}
{Onken}, C.~A., {Ferrarese}, L., {Merritt}, D., {Peterson}, B.~M., {Pogge},
  R.~W., {Vestergaard}, M., \& {Wandel}, A. 2004, \apj, 615, 645

\bibitem[{{Parry} {et~al.}(2009){Parry}, {Eke}, \& {Frenk}}]{parr09}
{Parry}, O.~H., {Eke}, V.~R., \& {Frenk}, C.~S. 2009, \mnras, 396, 1972

\bibitem[{Peng(2009)}]{peng09}
Peng, C.~Y. 2009, in prep.

\bibitem[{Peng {et~al.}(2002)Peng, Ho, Impey, \& Rix}]{peng02}
Peng, C.~Y., Ho, L.~C., Impey, C.~D., \& Rix, H.-W. 2002, AJ, 124, 266

\bibitem[{{Peng} {et~al.}(2006{\natexlab{a}}){Peng}, {Impey}, {Ho}, {Barton},
  \& {Rix}}]{peng06a}
{Peng}, C.~Y., {Impey}, C.~D., {Ho}, L.~C., {Barton}, E.~J., \& {Rix}, H.-W.
  2006{\natexlab{a}}, \apj, 640, 114

\bibitem[{{Peng} {et~al.}(2006{\natexlab{b}}){Peng}, {Impey}, {Rix},
  {Kochanek}, {Keeton}, {Falco}, {Leh{\'a}r}, \& {McLeod}}]{peng06b}
{Peng}, C.~Y., {Impey}, C.~D., {Rix}, H.-W., {Kochanek}, C.~S., {Keeton},
  C.~R., {Falco}, E.~E., {Leh{\'a}r}, J., \& {McLeod}, B.~A.
  2006{\natexlab{b}}, \apj, 649, 616

\bibitem[{{Riechers} {et~al.}(2008{\natexlab{a}}){Riechers}, {Walter},
  {Brewer}, {Carilli}, {Lewis}, {Bertoldi}, \& {Cox}}]{riec08a}
{Riechers}, D.~A., {Walter}, F., {Brewer}, B.~J., {Carilli}, C.~L., {Lewis},
  G.~F., {Bertoldi}, F., \& {Cox}, P. 2008{\natexlab{a}}, \apj, 686, 851

\bibitem[{{Riechers} {et~al.}(2008{\natexlab{b}}){Riechers}, {Walter},
  {Carilli}, {Bertoldi}, \& {Momjian}}]{riec08b}
{Riechers}, D.~A., {Walter}, F., {Carilli}, C.~L., {Bertoldi}, F., \&
  {Momjian}, E. 2008{\natexlab{b}}, \apjl, 686, L9

\bibitem[{{Riechers} {et~al.}(2009){Riechers}, {Walter}, {Carilli}, \&
  {Lewis}}]{riec09}
{Riechers}, D.~A., {Walter}, F., {Carilli}, C.~L., \& {Lewis}, G.~F. 2009,
  \apj, 690, 463

\bibitem[{Robaina {et~al.}(2009)Robaina, Bell, Skelton, E., McIntosh,
  Somerville, Zheng, Rix, Bacon, Balogh, Barazza, Barden, B\"ohm, Caldwell,
  Gallazzi, Gray, H\"aussler, Heymans, Jahnke, Jogee, van Kampen, Lane,
  Meisenheimer, Papovich, Peng, Sanchez, Skibba, Taylor, Wisotzki, \&
  Wolf}]{roba09}
Robaina, A.~R., Bell, E.~F., Skelton, E., R., McIntosh, D.~H., Somerville,
  R.~S., Zheng, X., Rix, H.-W., Bacon, D., Balogh, M., Barazza, F.~D., Barden,
  M., B\"ohm, A., Caldwell, J.~A.~R., Gallazzi, A., Gray, M.~E., H\"aussler,
  B., Heymans, C., Jahnke, K., Jogee, S., van Kampen, E., Lane, K.,
  Meisenheimer, K., Papovich, C., Peng, C.~Y., Sanchez, S., Skibba, R., Taylor,
  A., Wisotzki, L., \& Wolf, C. 2009, submitted to ApJ

\bibitem[{{Salvato} {et~al.}(2009){Salvato}, {Hasinger}, {Ilbert}, {Zamorani},
  {Brusa}, {Scoville}, {Rau}, {Capak}, {Arnouts}, {Aussel}, {Bolzonella},
  {Buongiorno}, {Cappelluti}, {Caputi}, {Civano}, {Cook}, {Elvis}, {Gilli},
  {Jahnke}, {Kartaltepe}, {Impey}, {Lamareille}, {LeFloch}, {Lilly},
  {Mainieri}, {McCarthy}, {McCracken}, {Mignoli}, {Mobasher}, {Murayama},
  {Sasaki}, {Sanders}, {Schiminovich}, {Shioya}, {Shopbell}, {Silverman},
  {Smol{\v c}i{\'c}}, {Surace}, {Taniguchi}, {Thompson}, {Trump}, {Urry}, \&
  {Zamojski}}]{salv09}
{Salvato}, M., {Hasinger}, G., {Ilbert}, O., {Zamorani}, G., {Brusa}, M.,
  {Scoville}, N.~Z., {Rau}, A., {Capak}, P., {Arnouts}, S., {Aussel}, H.,
  {Bolzonella}, M., {Buongiorno}, A., {Cappelluti}, N., {Caputi}, K., {Civano},
  F., {Cook}, R., {Elvis}, M., {Gilli}, R., {Jahnke}, K., {Kartaltepe}, J.~S.,
  {Impey}, C.~D., {Lamareille}, F., {LeFloch}, E., {Lilly}, S., {Mainieri}, V.,
  {McCarthy}, P., {McCracken}, H., {Mignoli}, M., {Mobasher}, B., {Murayama},
  T., {Sasaki}, S., {Sanders}, D.~B., {Schiminovich}, D., {Shioya}, Y.,
  {Shopbell}, P., {Silverman}, J., {Smol{\v c}i{\'c}}, V., {Surace}, J.,
  {Taniguchi}, Y., {Thompson}, D., {Trump}, J.~R., {Urry}, M., \& {Zamojski},
  M. 2009, \apj, 690, 1250

\bibitem[{S\'anchez {et~al.}(2004)S\'anchez, Jahnke, Wisotzki, Barden,
  Beckwith, Bell, Borch, Caldwell, H\"au{\ss}ler, Heymans, Jogee, McIntosh,
  Meisenheimer, Peng, Rix, Somerville, \& Wolf}]{sanc04a}
S\'anchez, S.~F., Jahnke, K., Wisotzki, L., Barden, M., Beckwith, S.~V.~W.,
  Bell, E.~F., Borch, A., Caldwell, J.~A.~R., H\"au{\ss}ler, B., Heymans, C.,
  Jogee, S., McIntosh, D.~H., Meisenheimer, K., Peng, C.~Y., Rix, H.-W.,
  Somerville, R.~S., \& Wolf, C. 2004, ApJ, 614, 586

\bibitem[{{Schramm} {et~al.}(2008){Schramm}, {Wisotzki}, \& {Jahnke}}]{schr08}
{Schramm}, M., {Wisotzki}, L., \& {Jahnke}, K. 2008, \aap, 478, 311

\bibitem[{{Scoville} {et~al.}(2007){Scoville}, {Abraham}, {Aussel}, {Barnes},
  {Benson}, {Blain}, {Calzetti}, {Comastri}, {Capak}, {Carilli}, {Carlstrom},
  {Carollo}, {Colbert}, {Daddi}, {Ellis}, {Elvis}, {Ewald}, {Fall},
  {Franceschini}, {Giavalisco}, {Green}, {Griffiths}, {Guzzo}, {Hasinger},
  {Impey}, {Kneib}, {Koda}, {Koekemoer}, {Lefevre}, {Lilly}, {Liu},
  {McCracken}, {Massey}, {Mellier}, {Miyazaki}, {Mobasher}, {Mould}, {Norman},
  {Refregier}, {Renzini}, {Rhodes}, {Rich}, {Sanders}, {Schiminovich},
  {Schinnerer}, {Scodeggio}, {Sheth}, {Shopbell}, {Taniguchi}, {Tyson}, {Urry},
  {Van Waerbeke}, {Vettolani}, {White}, \& {Yan}}]{scov07b}
{Scoville}, N., {Abraham}, R.~G., {Aussel}, H., {Barnes}, J.~E., {Benson}, A.,
  {Blain}, A.~W., {Calzetti}, D., {Comastri}, A., {Capak}, P., {Carilli}, C.,
  {Carlstrom}, J.~E., {Carollo}, C.~M., {Colbert}, J., {Daddi}, E., {Ellis},
  R.~S., {Elvis}, M., {Ewald}, S.~P., {Fall}, M., {Franceschini}, A.,
  {Giavalisco}, M., {Green}, W., {Griffiths}, R.~E., {Guzzo}, L., {Hasinger},
  G., {Impey}, C., {Kneib}, J.-P., {Koda}, J., {Koekemoer}, A., {Lefevre}, O.,
  {Lilly}, S., {Liu}, C.~T., {McCracken}, H.~J., {Massey}, R., {Mellier}, Y.,
  {Miyazaki}, S., {Mobasher}, B., {Mould}, J., {Norman}, C., {Refregier}, A.,
  {Renzini}, A., {Rhodes}, J., {Rich}, M., {Sanders}, D.~B., {Schiminovich},
  D., {Schinnerer}, E., {Scodeggio}, M., {Sheth}, K., {Shopbell}, P.~L.,
  {Taniguchi}, Y., {Tyson}, N.~D., {Urry}, C.~M., {Van Waerbeke}, L.,
  {Vettolani}, P., {White}, S.~D.~M., \& {Yan}, L. 2007, \apjs, 172, 38

\bibitem[{S\'ersic(1968)}]{sers68}
S\'ersic, J. 1968, Atlas de Galaxias Australes, Observatorio Astronomico de
  Cordoba

\bibitem[{{Simmons} \& {Urry}(2008)}]{simm08}
{Simmons}, B.~D., \& {Urry}, C.~M. 2008, \apj, 683, 644

\bibitem[{{Somerville} {et~al.}(2008){Somerville}, {Hopkins}, {Cox},
  {Robertson}, \& {Hernquist}}]{some08}
{Somerville}, R.~S., {Hopkins}, P.~F., {Cox}, T.~J., {Robertson}, B.~E., \&
  {Hernquist}, L. 2008, \mnras, 391, 481

\bibitem[{{Taylor} {et~al.}(2009){Taylor}, {Franx}, {van Dokkum}, {Bell},
  {Brammer}, {Rudnick}, {Wuyts}, {Gawiser}, {Lira}, {Urry}, \& {Rix}}]{tayl09}
{Taylor}, E.~N., {Franx}, M., {van Dokkum}, P.~G., {Bell}, E.~F., {Brammer},
  G.~B., {Rudnick}, G., {Wuyts}, S., {Gawiser}, E., {Lira}, P., {Urry}, C.~M.,
  \& {Rix}, H.-W. 2009, \apj, 694, 1171

\bibitem[{{Tremonti} {et~al.}(2004){Tremonti}, {Heckman}, {Kauffmann},
  {Brinchmann}, {Charlot}, {White}, {Seibert}, {Peng}, {Schlegel}, {Uomoto},
  {Fukugita}, \& {Brinkmann}}]{trem04}
{Tremonti}, C.~A., {Heckman}, T.~M., {Kauffmann}, G., {Brinchmann}, J.,
  {Charlot}, S., {White}, S.~D.~M., {Seibert}, M., {Peng}, E.~W., {Schlegel},
  D.~J., {Uomoto}, A., {Fukugita}, M., \& {Brinkmann}, J. 2004, \apj, 613, 898

\bibitem[{{Treu} {et~al.}(2007){Treu}, {Woo}, {Malkan}, \&
  {Blandford}}]{treu07}
{Treu}, T., {Woo}, J.-H., {Malkan}, M.~A., \& {Blandford}, R.~D. 2007, \apj,
  667, 117

\bibitem[{{Trump} {et~al.}(2009{\natexlab{a}}){Trump}, {Impey}, {Elvis},
  {McCarthy}, {Huchra}, {Brusa}, {Salvato}, {Capak}, {Cappelluti}, {Civano},
  {Comastri}, {Gabor}, {Hao}, {Hasinger}, {Jahnke}, {Kelly}, {Lilly},
  {Schinnerer}, {Scoville}, \& {Smol{\v c}i{\'c}}}]{trum09a}
{Trump}, J.~R., {Impey}, C.~D., {Elvis}, M., {McCarthy}, P.~J., {Huchra},
  J.~P., {Brusa}, M., {Salvato}, M., {Capak}, P., {Cappelluti}, N., {Civano},
  F., {Comastri}, A., {Gabor}, J., {Hao}, H., {Hasinger}, G., {Jahnke}, K.,
  {Kelly}, B.~C., {Lilly}, S.~J., {Schinnerer}, E., {Scoville}, N.~Z., \&
  {Smol{\v c}i{\'c}}, V. 2009{\natexlab{a}}, \apj, 696, 1195

\bibitem[{{Trump} {et~al.}(2009{\natexlab{b}}){Trump}, {Impey}, {Kelly},
  {Elvis}, {Merloni}, {Bongiorno}, {Gabor}, {Hao}, {McCarthy}, {Huchra},
  {Brusa}, {Cappelluti}, {Koekemoer}, {Nagao}, {Salvato}, \&
  {Scoville}}]{trum09b}
{Trump}, J.~R., {Impey}, C.~D., {Kelly}, B.~C., {Elvis}, M., {Merloni}, A.,
  {Bongiorno}, A., {Gabor}, J., {Hao}, H., {McCarthy}, P.~J., {Huchra}, J.~P.,
  {Brusa}, M., {Cappelluti}, N., {Koekemoer}, A., {Nagao}, T., {Salvato}, M.,
  \& {Scoville}, N.~Z. 2009{\natexlab{b}}, \apj, 700, 49

\bibitem[{{Walter} {et~al.}(2004){Walter}, {Carilli}, {Bertoldi}, {Menten},
  {Cox}, {Lo}, {Fan}, \& {Strauss}}]{walt04}
{Walter}, F., {Carilli}, C., {Bertoldi}, F., {Menten}, K., {Cox}, P., {Lo},
  K.~Y., {Fan}, X., \& {Strauss}, M.~A. 2004, \apj, 615, L17

\bibitem[{{Woo} {et~al.}(2008){Woo}, {Treu}, {Malkan}, \& {Blandford}}]{woo08}
{Woo}, J.-H., {Treu}, T., {Malkan}, M.~A., \& {Blandford}, R.~D. 2008, \apj,
  681, 925

\bibitem[{Zheng {et~al.}(2009)Zheng, Bell, Somerville, Rix, Jahnke, Fontanot,
  Rieke, Schiminovich, \& Meisenheimer}]{zhen09}
Zheng, X.~Z., Bell, E.~F., Somerville, R.~S., Rix, H.-W., Jahnke, K., Fontanot,
  F., Rieke, G.~H., Schiminovich, D., \& Meisenheimer, K. 2009, submitted to
  ApJ

\end{thebibliography}

\end{document}